\documentstyle[prl,floats,aps,twocolumn,epsf]{revtex}

\tighten

\begin{document}
\draft

\twocolumn[\hsize\textwidth\columnwidth\hsize\csname
@twocolumnfalse\endcsname

\rightline{
\large\baselineskip20pt\rm\vbox to20pt{
\baselineskip14pt
\hbox{OU-TAP-167}}}

\title{Calculating the gravitational self force in Schwarzschild spacetime}
\author{Leor Barack$^1$, Yasushi Mino$^2$, Hiroyuki Nakano$^3$, Amos Ori$^4$,
and Misao Sasaki$^3$}
\address{
$^1$Albert-Einstein-Institut, Max-Planck-Institut f{\"u}r Gravitationsphysik,
Am M\"uhlenberg 1, D-14476 Golm, Germany\\
$^2$Theoretical Astrophysics, California institute of Technology, Pasadena,
California 91125\\
$^3$Department of Earth and Space Science, Graduate School of Science,
Osaka University, Toyonaka, Osaka 560-0043, Japan\\
$^4$Department of Physics, Technion---Israel Institute of Technology,
Haifa, 32000, Israel}
\date{\today}
\maketitle

\begin{abstract}
We present a practical method for calculating the local gravitational
self-force (often called ``radiation-reaction force'') for a pointlike
particle orbiting a Schwarzschild black hole. This is an implementation
of the method of {\it mode-sum regularization}, in which one first
calculates the (finite) contribution to the
force due to each individual multipole mode of the perturbation,
and then applies a certain regularization procedure to the mode sum.
Here we give the values of all the ``regularization parameters'' (RP)
required for implementing this regularization procedure, for any
geodesic orbit in Schwarzschild spacetime.

\end{abstract}


\vspace{2ex}
]


The problem of determining the self force (SF) acting on a charged particle
dates many decades back to the classical works by Lorentz \cite{Lorentz} and
Dirac \cite{Dirac} on the electron's equation of motion. This force---often
referred to as ``radiation reaction force''---may be attributed to the flux
of energy-momentum carried away by the electromagnetic radiation. Recent
years' growing interest in sources of gravitational waves has renewed
interest in this old problem, in a new and exciting context: the
{\it gravitational} SF acting on a point mass in curved spacetime.

This manuscript deals with the motion of a small pointlike particle of mass
$\mu $ around a Schwarzschild black hole with mass $M\gg\mu $. When $\mu$
is finite, the particle's trajectory deviates from a geodesic of the
Schwarzschild background geometry, due to the particle's interaction with
its own gravitational field. This deviation indicates the presence of a
self force. The particle's equation of motion thus takes the form
$\mu a_{\alpha}=F_{\alpha}^{\rm self}$, where $a_{\alpha}$ is the
covariant acceleration (i.e., $a_{\alpha}=u_{\alpha;\beta}u^{\beta}$),
and $F_{\alpha}^{\rm self}\propto O(\mu^2)$ describes the leading-order
gravitational SF effect. In general, knowledge of $F_{\alpha}^{\rm self}$
is essential for modeling the orbital evolution in binary black-hole systems
with extreme mass ratios. Such astrophysical systems---typically, the scenario
of a small compact object captured by a supermassive black hole (of the kind
now believed to reside in the cores of many galaxies, including our own)---are
expected to serve as main targets for future space-based gravitational wave
detectors \cite{LISA}. The need for precisely predicting the orbital
evolution of such binary systems---and the waveform of the emitted
gravitational radiation---strongly motivates the development of practical
methods for gravitational SF calculations.

When dealing with point particles, one faces the fundamental issue of
{\em regularization}: extracting the ``correct'' finite part of the
(otherwise divergent) SF. A formalism for regularizing the electromagnetic
SF in curved spacetime was developed long ago by DeWitt and Brehme \cite{DB}.
Recently, Mino, Sasaki, and Tanaka (MST)\ \cite{MST} first obtained a formal
expression for the {\em gravitational} SF, based on a physically
well-motivated regularization technique. The same formal result was
independently derived by Quinn and Wald (QW) \cite{QW} using an axiomatic
approach. It is useful to write this result as
\begin{equation} \label{Fself}
F_{\alpha}^{{\rm self}}=
\lim_{x\to z_0}\left[F_{\alpha}^{{\rm full}}(x)-
F_{\alpha}^{{\rm dir}}(x)\right],
\end{equation}
where $z_0^{\mu}$ is the SF's evaluation point, $x^{\mu}$ is a point in the
neighborhood of $z_0^{\mu}$, $F_{\alpha}^{{\rm full}}$ is the ``full'' force
constructed from the metric perturbation (as described below), and
$F_{\alpha}^{{\rm dir}}$ (the ``direct'' force) is
the ``divergent piece'' to be removed, whose formal construction is
described in \cite{MST,QW}. Roughly speaking, the direct piece
$F_{{\rm dir}}^{\alpha}$ represents the instantaneous effect of waves
propagating
directly along the particle's light cone; hence, the SF effect is entirely
attributed to the so-called ``tail'' part of the particle's field, i.e.,
waves scattered {\em inside} the particle's past light cone due to the
background's curvature.

The above regularization prescription by MST is based on the physically
well-motivated method of matched asymptotic expansions. At present, this
prescription---which also conforms with QW's prescription---is widely
accepted as the standard regularization scheme for the SF
\cite{alternative}. As it stands,
however, it is difficult to directly implement this method in practical
calculations, because it requires the computation of the Green's function
from any worldline point to any other future worldline point \cite{Weakfield}.
To allow the practical implementation of Eq.\ (\ref{Fself}) in strong-field
calculations, Barack and Ori (BO) devised a method based on a mode
decomposition of the perturbation field (see \cite{MSRS-scalar} for the
scalar SF and \cite{MSRS-grav} for the gravitational SF).
In the mode-sum method, the quantities
$F_{\alpha}^{{\rm full}}$ and $F_{\alpha}^{{\rm dir}}$ are first formally
expanded into multipole $l$-modes $F_{\alpha l}^{{\rm full}}$ and
$F_{\alpha l}^{{\rm dir}} $. Most benificial is the fact that, whereas
$F_{\alpha}^{{\rm full}}$ diverges as $x\to z_0$, its individual modes
attain finite values even at the particle's location. The full mode
$F_{\alpha l}^{{\rm full}}$ is directly derived (see below) from the
$l$-mode perturbation field, which, in turn, is obtained from the
decoupled field equations \cite{MSRS-grav,Perturbation} by standard
numerical techniques. [The construction of $F_{\alpha l}^{{\rm full}}$
involves summation over the azimuthal number $m$, and---in the
gravitational SF case---also over the ten tensorial harmonics for
each multipole number $l$.] The direct piece
$F_{\alpha}^{{\rm dir}}$ (and, hence, its $l$ mode
$F_{\alpha l}^{{\rm dir}}$) is analyzed by local methods (see below).
The expression (\ref{Fself}) for the SF is then brought to the form
\cite {MSRS-scalar,MSRS-grav}
\begin{equation}\label{MSRS}
F_{\alpha}^{{\rm self}}=\sum_{l=0}^{\infty}\left[ \lim_{x\to z_0}F_{\alpha
l}^{{\rm full}}-A_{\alpha}L-B_{\alpha}-C_{\alpha}/L\right] -D_{\alpha},
\end{equation}
where hereafter the limit $x\to z_0$ is taken along the (ingoing or outgoing)
radial direction, $L\equiv l+1/2$, and $A_{\alpha}$, $B_{\alpha}$,
$C_{\alpha}$, and $D_{\alpha}$ are certain ``regularization parameters''
(RP), derived from the local structure of $F_{\alpha l}^{{\rm dir}}$: Roughly
speaking, $A_{\alpha}L+B_{\alpha}+C_{\alpha}/L$ reflects the asymptotic
form of $F_{\alpha l}^{{\rm dir}}$ (and also $F_{\alpha l}^{{\rm full}}$) at
large $l$, while $D_{\alpha}$ is a residual quantity that arises in the
summation over $l$ (see \cite{MSRS-scalar,MSRS-grav} for a more precise
definition of the RP). Note that in the framework of the mode-sum method
the task of calculating the SF is divided into two separate parts:
(i) Calculating the quantities $F_{\alpha l}^{{\rm full}}$---e.g., by
numerically integrating the decomposed field equations; and
(ii) analytically calculating the four RP, by locally analyzing the direct
part \cite{MSRS-implement}.
It is the second part, the analytic derivation of the RP, that will
primarily concern us in this Letter.

In the original method by BO, the analysis of $F_{\alpha l}^{{\rm dir}}$
was carried out by applying a systematic perturbation expansion of the
($l$-mode) Green's equation. The derivation of the full set of four RP
required the calculation of the Green function up to second order in this
expansion. So far, the RP where calculated using this original
method in a few special cases: the scalar SF for a particle in radial or
circular orbits \cite{MSRS-scalar}, and the gravitational SF for radial
motion \cite{MSRS-grav,implementation} (all in Schwarzschild spacetime).
However, the required second-order perturbation analysis of the Green
function turned out to be rather cumbersome, making it difficult to
extend the calculation to more general cases.

The goal of this Letter is to report on a new method for analytically
deriving the four RP, which proves to be significantly simpler than the
original BO's method. This new mode-sum method is based on an analytic
expression constructed by Mino, Nakano and Sasaki [MNS]
\cite{MinoNakano,direct} for the direct force $F_{\alpha}^{\rm dir}$.
(MNS developed this expression as part of a more comprehensive study
aimed to provide a systematic description of the local
structure of the metric perturbation.)
Once this expression
is available, essentially all we need is to decompose it into $l$-modes,
and then take the limit $x\to z_0$. The large-$l$ asymptotic behavior of
this expansion turns out to be a power series in $1/L$, starting at
$L^{1}$, and its three leading-order coefficients yield the three RP
$A_{\alpha},B_{\alpha},C_{\alpha}$ according to
\begin{equation}\label{ABC}
\lim_{x\to z_0}F_{\alpha l}^{\rm dir}=A_{\alpha}L+B_{\alpha}
+C_{\alpha}/L+O(L^{-2}).
\end{equation}
The last parameter $D_{\alpha}$ is then obtained by
\begin{equation}\label{D}
D_{\alpha}=
\sum_{l=0}^{\infty}\left[\lim_{x\to z_0}F_{\alpha l}^{\rm dir}
-A_{\alpha}L-B_{\alpha}-C_{\alpha}/L\right].
\end{equation}
Decomposing the right-hand side of Eq.\ (\ref{Fself}) and then substituting
Eq.\ (\ref{D}), one recovers Eq.\ (\ref{MSRS}). (Note that, since the tail
part is regular, $F_{\alpha l}^{{\rm full}}$ and $F_{\alpha l}^{{\rm dir}}$
admit the same large-$l$ asymptotic behavior, with the same coefficients.)

Applying this new method, we have been able to derive the values of all RP
for an arbitrary geodesic orbit in Schwarzschild spacetime, for both the
scalar and gravitational self-force, as we now describe. Throughout this
paper we use geometrized units $G=c=1$ and metric signature ${-}{+}{+}{+}$,
and denote the standard Schwarzschild coordinates by $t,r,\theta,\varphi$.

We begin by considering a simple scalar-field toy model. Once the scalar
case is well understood, it becomes fairly straightforward to obtain the RP
in the gravitational case as well. We shall consider a particle carrying a
scalar charge $q$, which (in the lack of SF) moves along an equatorial geodesic
$z^{\mu}(\tau)$ of the Schwarzschild background ($\tau$ being the proper
time) with specific energy and angular momentum $\cal E$ and $\cal L$,
respectively. The scalar field $\Phi$ satisfies the field equation
$\Phi_{;\alpha}^{\; ;\alpha}=-4\pi\rho$,
where $\rho=q\int \delta^4(x-z)(-g)^{-1/2}d\tau$ is
the scalar charge density ($g$ being the metric determinant).
In this model, the force exerted on the scalar charge by an (external, regular)
scalar field $\Phi(x)$ is simply given by $F_{\alpha}=q\Phi_{,\alpha}$.

Let $\epsilon (x)$ denote the spatial distance from $x^{\mu}$ to the
worldline $z^{\mu}(\tau)$ (i.e., the length of the short geodesic
connecting $x^{\mu}$ to the worldline and normal to it). Using the Hadamard
expansion of the field equation, MNS have been able to show that the
``direct'' scalar force takes the form \cite{direct}
\begin{equation}\label{Fdir0}
F_{\alpha}^{{\rm dir}}(x)=q\Phi_{,\alpha}^{{\rm dir}},\quad\quad
\Phi^{\rm dir}(x)\equiv q\hat{f}/\epsilon,
\end{equation}
where $\Phi^{{\rm dir}}(x)$ is the so called ``direct scalar field'', and
$\hat{f}(x)$ is some regular scalar function satisfying
$\hat{f}=1+O(\delta x^{2})$, where $\delta x^{\mu}\equiv x^{\mu}-z_0^{\mu}$
(the explicit form of $\hat{f}$ will not be needed here). For the analysis
below it is useful to define $S\equiv\epsilon ^{2}$. The direct force then takes
the form
\begin{equation}\label{Fdir}
F_{\alpha}^{\rm dir}(x)=
q^2\left(S^{-1/2}\hat{f}_{,\alpha}-\hat{f}S^{-3/2}S_{,\alpha}/2\right).
\end{equation}
Being a regular function of $x$, $S$ may be expanded around $z_0$ as
$S=S_{0}+O(\delta x^{3})$, where \cite{details} $S_{0}=(g_{\mu\nu}+
u_{\mu}u_{\nu})\delta x^{\mu}\delta x^{\nu}$, and $u^{\mu}\equiv
dz^{\mu}/d\tau $. (In the expression for $S_{0}$, $g_{\mu \nu}$ and
$u_{\alpha}$ are to be evaluated at $z_0$.)

In the next step we need to decompose the components $F_{\alpha}^{{\rm dir}}$
into spherical harmonics and then derive the four RP, as prescribed above.
This goal was undertaken independently (and in two slightly different ways)
by our two groups, BO and MNS, as we now describe. \cite{Burko}

In the analysis by BO \cite{details}, the expression $S=S_{0}+O(\delta
x^{3}) $ is first substituted in Eq.\ (\ref{Fdir}). With some simple
manipulations, this equation can be brought to the form
\begin{equation}\label{expansion}
F_{\alpha}^{{\rm dir}}(x)=\epsilon _{0}^{-3}P_{\alpha}^{(1)}+
\epsilon_{0}^{-5}P_{\alpha}^{(4)}+\epsilon_{0}^{-7}P_{\alpha}^{(7)}
+O(\delta x),
\end{equation}
where $\epsilon_{0}\equiv S_{0}^{1/2}$, and $P_{\alpha}^{(n)}$ denote terms of
homogeneous order $n$ in $\delta x^{\alpha}$. Notice that the three terms
on the right hand side are of order $\delta x^{-2}$, $\delta x^{-1}$,
and $\delta x^{0}$, respectively. The $O(\delta x)$ correction term will not
concern us here, as it does not contribute to the direct force at $z_0$.

As explained above, the mode $l$ of $F_{\alpha}^{{\rm dir}}$ (or
$F_{\alpha}^{{\rm full}}$) is obtained by summing the contributions of all
possible values of $m$ for this specific $l$. This sum is invariant under
rotation of the angular variables $\theta,\varphi $. Using this invariance,
BO chose a new, rotated, set of angular variables $\theta',\varphi'$
(with respect to which the spherical harmonics are to be defined), such that
the evaluation point $z_0$ is located at the pole, $\theta'=0$.
Since all $m\neq 0$ spherical harmonics vanish at $\theta'=0$,
with these rotated coordinates one only needs to consider the $m=0$ modes.
This greatly simplifies the analysis. The presence of a coordinate
singularity (the trivial $\theta'=0$ singularity) at the
evaluation point somewhat complicates the situation. To overcome this
difficulty, BO introduced two regular ``Cartesian-like'' coordinates $x,y$.
With this choice of coordinates $(t,r,x,y)$, the $l$-decomposition of Eq.\
(\ref{expansion}) becomes especially simple.
After taking the limit $x\to z_0$, one finds that
(i) the term $\epsilon_{0}^{-7}P_{\alpha}^{(7)}$ yields {\em zero
contribution} to $F_{\alpha l}^{\rm dir}$;
(ii) the term $\epsilon_{0}^{-5}P_{\alpha}^{(4)}$ yields a {\em constant}
(i.e., $l$-independent) contribution, which we denote $b_{\alpha}$; and
(iii) the contribution of the term $\epsilon_{0}^{-3}P_{\alpha}^{(1)}$ is
{\em precisely} proportional to $L$---we shall denote it $a_{\alpha}L$.
From Eqs. (\ref{ABC}) and (\ref{D}) it now follows that $A_{\alpha}=a_{\alpha}$,
$B_{\alpha}=b_{\alpha}$, and $C_{\alpha}=D_{\alpha}=0$. The detailed
calculation of the coefficients $a_{\alpha}$ and $b_{\alpha}$ is given in Ref.\
\cite{details}. The final expressions obtained by BO for the RP (expressed in the
original $\theta ,\varphi $ coordinates, in which the motion is equatorial) are
given by $A_{\theta}^{\rm sc}=B_{\theta}^{{\rm sc}}=0$ ,
\begin{mathletters}\label{RPscalar}
\begin{equation} \label{A}
A_{\pm r}^{\rm sc}=\mp \frac{q^{2}}{r^{2}}\,\frac{\cal E}{fV},
\quad
A_{\pm t}^{\rm sc}=\pm \frac{q^{2}}{r^{2}}\,\frac{\dot{r}}{V},
\quad A_{\varphi}^{\rm sc}=0,
\end{equation}
\begin{equation} \label{Br}
B_{r}^{\rm sc}=\frac{q^2}{r^2}\,\frac{(\dot{r}^2-2{\cal E}^2)\hat{K}(w)
+(\dot{r}^2+{\cal E}^2)\hat{E}(w)}{\pi fV^{3/2}},
\end{equation}
\begin{equation} \label{Bt}
B_{t}^{\rm sc}=\frac{q^2}{r^{2}}\,\frac{{\cal E}\dot{r}
[\hat{K}(w)-2\hat{E}(w)]}{\pi V^{3/2}},
\end{equation}
\begin{equation}\label{Bphi}
B_{\varphi}^{\rm sc}=\frac{q^2}{r}\,\frac{\dot{r}[\hat{K}(w)-\hat{E}(w)]}
{\pi ({\cal L}/r)V^{1/2}},
\end{equation}
\begin{equation}\label{CD}
C_{\alpha}^{\rm sc}=D_{\alpha}^{\rm sc}=0
\end{equation}
\end{mathletters}
(with `sc' signifying the {\em scalar} force RP), where $\hat{K}(w)$ and
$\hat{E}(w)$ are the complete elliptic integrals of the first and second
kinds, respectively, $w\equiv {\cal L}^2/({\cal L}^2+r^2)$,
$f\equiv (1-2M/r)$, $V\equiv 1+{\cal L}^2/r^2$, and $\dot{r}\equiv dr/d\tau$.
The ``$\pm $'' sign in $A_{\alpha}$ refers to whether the limit $x\to z_0$ is
taken along the ingoing or outgoing radial direction. The values
(\ref{RPscalar}) agree with those obtained in \cite{MSRS-scalar}
in the special cases ${\cal L}=0$ or $\dot{r}=0$.

In an independent analysis, MNS directly used the ``standard'' $\theta,\varphi$
coordinates (in which the motion is equatorial) for decomposing the direct
field. In this setup the $m\neq 0$ modes contribute as well. MNS derived an
analytic expression for the contribution of each $l,m$ mode of the direct
force, expanded in powers of $M/r$. Then, MNS have shown the convergence of this
expansion, and were able to explicitly sum it up (and sum over $m$), after which
they recovered all RP values (\ref{RPscalar}).


We now turn to discuss the gravitational SF. Let
$\bar{h}_{\alpha\beta}^{{\rm full}}\equiv \bar{h}_{\alpha\beta}$ denote
the trace-reversed metric perturbation induced by a particle of mass $\mu$
(namely, $\bar{h}_{\alpha\beta}\equiv h_{\alpha\beta}-\frac{1}{2}g_{\alpha\beta}
g^{\mu\nu}h_{\mu\nu}$, where $h_{\alpha\beta}$ is the metric perturbation
itself and $g_{\alpha\beta}$ is the background metric).
Just as in the scalar case, this ``full'' perturbation is the sum of
the ``direct'' part and the ``tail'' part, $\bar{h}_{\alpha \beta}
^{\rm full}=\bar{h}_{\alpha\beta}^{\rm dir}+\bar{h}_{\alpha \beta}
^{\rm tail}$. The gravitational SF is then obtained by applying a certain
differential operator to $\bar{h}_{\alpha \beta}^{\rm tail}$:
\begin{equation} \label{gself}
F_{\alpha}^{\rm self}=F_{\alpha}^{{\rm tail}}\equiv
\mu\, k_{\alpha}{}^{\beta\gamma \delta}\bar{h}_{\beta\gamma;\delta}^{{\rm tail}}
\end{equation}
evaluated at $x\to z_0$, where
\begin{eqnarray} \label{k}
k^{\alpha\beta\gamma\delta} &=&u^{\beta}u^{\gamma}g^{\alpha \delta}/2
+g^{\beta \gamma}g^{\alpha \delta}/4+u^{\alpha}g^{\beta\gamma}u^{\delta}/4
-g^{\alpha \beta}u^{\gamma}u^{\delta}  \nonumber \\
&&-u^{\alpha}u^{\beta}u^{\gamma}u^{\delta}/2,
\end{eqnarray}
all quantities evaluated at $z_0$. Let us also define
\begin{equation} \label{full-dir}
F_{\alpha}^{\rm (dir,full)}=\mu\, k_{\alpha}{}^{\beta\gamma\delta}
\bar{h}_{\beta\gamma;\delta}^{\rm (dir,full)},
\end{equation}
after which Eq.\ (\ref{Fself}) is recovered.

The direct trace-reversed metric perturbation was found by MST \cite{MST}.
It can be expressed as
\begin{equation} \label{hdir}
\bar{h}_{\beta\gamma}^{\rm dir}(x)=4\mu \bar{f}(x,z)\epsilon^{-1}
\hat{u}_{\beta}\hat{u}_{\gamma},
\end{equation}
where $\bar{f}$ is a regular function satisfying
$\bar{f}=1+O(\delta x^{2})$ and $\hat{u}_{\alpha}$ is the four-velocity
parallelly-propagated from the particle's worldline to $x$ along the
short normal geodesic.

As it stands, the tensor $k$ is only defined on the particle's worldline,
because it involves $u^{\alpha}$. In Eq.\ (\ref{gself}) this on-worldline
definition suffices, because $\bar{h}^{\rm tail}$ is regular at the
particle's location \cite{MST}. However, $\bar{h}^{{\rm full}}$ and
$\bar{h}^{\rm dir}$ diverge like $1/\epsilon $ (and their gradients like
$\epsilon^{-2}$). Therefore, when using Eq.\ (\ref{full-dir}) [e.g., when
substituting it in Eq.\ (\ref{Fself})] we must prescribe an extension
of $k$ {\em off} the worldline. Recall, however, that no ambiguity is caused
by this non-uniqueness of $k$: One just needs to use {\em the same extension}
for both the direct and full forces. Below we consider several extensions:
The one obtained by
substituting $u^{\alpha}\to \hat u^{\alpha}$ is denoted $\hat{k}$; and the
extension based on ``fixed (contravariant) components'' (in the Schwarzschild
coordinates) is denoted $\tilde{k}$. A third extension, mentioned below, is
denoted $\bar{k}$. Correspondingly, we shall denote the quantities associated
with the above three extensions (e.g. the RP) by a hat, tilde, or an over-bar,
respectively.

The gravitational RP are obtained by decomposing the components
$F_{\alpha}^{{\rm dir}}$ into scalar spherical harmonics, just as in the
scalar case. In the $\hat k$-extension, MNS obtained \cite{direct}
$\hat{R}_{\alpha}^{\rm gr}=R_{\alpha}^{\rm sc}$ (with $q\to\mu$), where
$R_{\alpha}$ stands for all four RP, and the label ``gr'' signifies the
gravitational case values. In the $\tilde k$-extension, BO found \cite{details}
$\tilde{R}_{\alpha}^{\rm gr}=K_{\alpha}^{\lambda}R_{\lambda}^{\rm sc}$,
where $K_{\alpha}^{\lambda}\equiv \delta_{\alpha}^{\lambda}
+u_{\alpha}u^{\lambda}$.

For implementing the mode-sum method, Eq.\ (\ref{MSRS}), we must have at hand
$F_{\alpha l}^{{\rm full}}$, the {\em spherical harmonic} $l$ mode of the
($\alpha$ component of the) full force. In the scalar case, this quantity is
obtained by numerically computing the full-field mode $\Phi_l$ and
differentiating it with respect to $x^{\alpha}$. In the gravitational case
the procedure is a bit more involved. The linearized Einstein equation are
separated by the decomposition
\begin{equation} \label{Yilm}
\bar{h}_{\alpha\beta}=\sum_{lm}\sum_{i=1}^{10}
\bar{h}^{(i)lm}(r,t)\,Y^{(i)lm}_{\alpha\beta}(\theta,\varphi),
\end{equation}
where $Y^{(i)lm}_{\alpha\beta}$ are the ten (Zerilli-type) {\em tensorial}
harmonics \cite{MSRS-grav}. The quantities $\bar h^{(i)lm}$ can be computed
by numerically integrating the decoupled field equations. Then,
\begin{equation} \label{Fbare-grav}
F_{\alpha}^{\rm full}=\mu \sum_{ilm}k_{\alpha}{}^{\beta\gamma\delta}
\left[\bar{h}^{(i)lm}(r,t)\,Y^{(i)lm}_{\beta\gamma}(\theta,\varphi)
\right]_{;\delta}.
\end{equation}
To obtain $F_{\alpha l}^{\rm full}$, we need to decompose this quantity in
ordinary (i.e., {\em scalar}) spherical harmonics $Y^{lm}(\theta,\varphi)$,
and to sum over $i$ and $m$. The outcome of this decomposition will depend
on the extension of $k$.
This decomposition appears to be difficult to implement with the
$\hat k$-extension. For most extensions (including $\tilde{k}$), infinite
number of tensorial harmonics $\bar{h}^{(i)l'm}$
will contribute to a single scalar-harmonic term $F_{\alpha lm}^{{\rm full}}$
(due to the nontrivial dependence of $k$ on $\theta $). However, BO designed
a third extension $\bar{k}=\tilde{k}+\delta k$, where $\delta k$ is a
certain $O(\delta x^{2})$ correction term (see \cite{details}), in which
only finite number ($l-3\leq l'\leq l+3$) of tensorial harmonics
$\bar{h}^{(i)l'm}$ contribute to a single term
$F_{\alpha lm}^{\rm full}$. Then $F_{\alpha l}^{\rm full}$ is given by
\begin{equation}\label{full-l}
F_{\alpha l}^{\rm full}=\mu \sum_{i=1}^{10}\sum_{l'=l-3}^{l+3}
\sum_{m=-l'}^{l'}
\left[{\cal D}_{\alpha l}^{(i)l'm}\bar{h}^{(i)l'm}(r,t)\right]Y^{lm},
\end{equation}
where ${\cal D}_{\alpha l}^{(i)l'm}$ is a certain 1st-order differential
operator (independent of $\theta,\varphi$) given explicitly in \cite{details}.
It can be shown \cite{details} that the extension difference $\delta k$,
being $O(\delta x^2)$, does not modify the RP values; hence,
\begin{equation} \label{RPgrav}
\bar{R}_{\alpha}^{\rm gr}=\tilde{R}_{\alpha}^{{\rm gr}}=
K_{\alpha}^{\lambda}R_{\lambda}^{{\rm sc}}.
\end{equation}
These values coincide, for ${\cal L}=0$, with the ones obtained
\cite{implementation} using local analysis of the Green function.

Let us now summarize the prescription for constructing the gravitational SF:
(i) Numerically compute the radial functions
$\bar{h}^{(i)l'm}(r,t)$ (e.g., in the harmonic gauge
\cite{MSRS-grav});
(ii) use Eq.\ (\ref{full-l}) to construct the full modes
$F_{\alpha l}^{{\rm full}}$;
(iii) use Eq.\ (\ref{RPgrav}) [along with Eq.\ (\ref{RPscalar}), with
$q\to\mu$] to obtain the RP; and
(iv) apply Eq.\ (\ref{MSRS}).
This prescription is now being implemented by Barack and Lousto
\cite{implementation}.

Finally, it should be reminded that the gravitational SF is a
gauge-dependent notion, as discussed in Ref.\ \cite{gauge}.
Nevertheless, the RP are gauge-independent \cite{gauge}.
The above prescription applies to the SF associated with the
harmonic gauge, or any other gauge related to it by a regular gauge
transformation.


We would like to thank Lior Burko for interesting discussions and
stimulating interaction.
L.B.\ was supported by a Marie Curie Fellowship of the European Community
program IHP-MCIF-99-1 under contract number HPMF-CT-2000-00851.
Y.M.\ was supported by NSF Grant PHY-0099568, PHY-0096522 and
NASA Grant NAG5-10707.
H.N.\ and M.S.\ were supported in part by a Monbusho Grant-in-Aid
for Creative Research (No.~09NP0801), and by a Monbusho Grant-in-Aid for
Scientific Research (No.~12640269).
H.N.\ was also supported by Research Fellowships of Japan Society for the
Promotion of Science for Young Scientists, No.~2397.


\end{document}